\documentclass[12pt]{article}
\begin{document}
\thispagestyle{empty}
\begin{flushright}
hep-ph/0307096 \\
\end{flushright}
\vspace*{0.4in}
\begin{center}
{\Large\sc Single Photon Signals for Warped Quantum Gravity
at a Linear ${\bf e}^{\bf +} {\bf e}^{\bf -}$ Collider}
\\
\vspace*{0.4in}
{\large\sl Santosh Kumar Rai}
and 
{\large\sl Sreerup Raychaudhuri}
\\
\vspace*{0.2in}
Department of Physics, Indian Institute of Technology, Kanpur 208 016, India 
\\
E-mail: {\sf skrai@iitk.ac.in, sreerup@iitk.ac.in} \\
\vspace*{1.5in}
{\Large\sc Abstract}
\end{center}
\begin{quotation}
\noindent
We study the `single photon' process $e^+ e^- \to \gamma \nu \bar{\nu}$
with contributions due to exchange of massive gravitons in the
Randall-Sundrum model of low-scale quantum gravity. It is shown that for
significant regions in the parameter space, this process unambiguously
highlights the resonance structure of the graviton sector. Even in the
non-resonant part of the parameter space, we show that comparison with the
benchmark process $e^+ e^- \to \mu^+ \mu^-$ can clearly distinguish
signals for warped gravity from similar signals for large extra
dimensions.
\end{quotation}
\vfill
\newpage
\section{Introduction}

One of the most exciting theoretical developments of recent years has been
the idea that there could be one or more extra spatial dimensions and that
the observable Universe could be confined to a four-dimensional
hyper-surface in a higher dimensional bulk spacetime \cite{Akama}. Such
ideas, which fit in naturally with current ideas in superstring theory,
give rise to elegant solutions to the well-known gauge hierarchy problem
of high energy physics. What is even more interesting, perhaps, is the
suggestion that there could be observable signals of quantum gravity at
current and future accelerator experiments.

This relatively new body of ideas, commonly dubbed `Brane World
Phenomenology', bases itself on two main ideas: the concept of hidden
compact dimensions~\cite{Kaluza} and the string-theoretic idea of
$D$-branes~\cite{Polchinski}. There are two main scenarios, each having
variants. One is the so-called Arkani-Hamed-Dimopoulos-Dvali (ADD)  
model~\cite{ADD}, in which there are $d$ more extra spatial dimensions,
compactified on a $d$-torus of radius $R_c$ each way, which, together with
the four canonical Minkowski dimensions, constitutes the `bulk' spacetime.
In this scenario $R_c$ can be~\cite{Long} as large as 200~$\mu$m. However,
the Standard Model (SM) fields are confined to a four-dimensional slice of
spacetime, with thickness not more than $10^{-12}$~$\mu$m, which is dubbed
the `brane'. If the model is embedded in a string-theoretic framework, the
`brane' is, in fact a $D_3$-brane, i.e. a 3+1 dimensional hyper-surface on
which the ends of open strings are confined\footnote{It is not absolutely
essential to embed the model in a string theory, and the word `brane' or
`wall' is then used simply to denote a hyper-surface or domain wall where
SM fields are confined.}.  However, gravity, which is a property of
spacetime itself, must be free to propagate in the bulk. As a result

\begin{itemize} 
\item Planck's constant in the bulk $\hat{M}_P$ is related to Planck's
constant on the brane $M_P$($\simeq~1.2\times 10^{19}$~GeV) by
$\hat{M}_P^{2+d} ~R_c^d = (4\pi)^{d/2} ~\Gamma(d/2) ~M_P^2$ which means
that for $R_c \sim 200$~$\mu$m, it is possible to have $\hat{M}_P$ as low
as a TeV. This solves the gauge hierarchy problem simply by bringing down
the scale of new physics (i.e. gravity)  to about a TeV.  
\item There are a huge number of massive Kaluza-Klein excitations of the
(bulk) graviton field, as perceived on the brane. These collectively
produce effects of electroweak strength, which may be observable at
current experiments and those planned in the near future~\cite{Giudice,
Lykken,Kubyshin}.  
\end{itemize}

A major drawback of the ADD model is that it creates a new hierarchy
between the `string scale' $\hat{M}_P \sim 1$~TeV and the size of the
extra dimensions $R_c^{-1} \sim 1$~$\mu$eV. Moreover, it can be argued
that the huge (compared with the Planck length) size of the extra
dimensions is unstable under quantum corrections, which tend to shrink it
down until $\hat{M}_P \sim R_c^{-1} \sim M_P$. This problem is brilliantly
solved in the Randall--Sundrum (RS) model~\cite{RS}, which has grown out
of the ADD model.

In the RS model, there are {\it two} branes --- a `visible' brane
containing the SM fields and an `invisible' brane where gravity is strong
(as strong as the electroweak interaction) --- embedded in a
five-dimensional bulk, where the single extra dimension is a
$\mathbf{S}^1/\mathbf{Z}_2$ orbifold. i.e. a circle folded about a
diameter. Placing the two branes at the two orbifold fixed points $\phi =
0, \pi$, and assuming they have equal and opposite energy densities (brane
tensions), which are related to a negative cosmological constant in the
bulk, one obtains a `warped' solution to the five-dimensional Einstein
equations of the form
\begin{equation}
ds^2 = e^{-{\cal K}R_c\phi} g_{\mu\nu}dx^\mu dx^\nu + R_c^2 d\phi^2
\end{equation}
where ${\cal K}$ is the curvature of the fifth dimension. Gravity is
strong on the invisible brane at $\phi = 0$ and weak on the visible brane
$\phi = \pi$. It is now possible to choose all the fundamental energy
scales in the problem in the ballpark of the Planck scale provided ${\cal
K}R_c \simeq 12$. This moderate value reduces the `warp factor' $e^{-{\cal
K}R_c\pi}$ to about $10^{-16}$ --- adequate to explain the hierarchy
between electroweak and Planck scales. One thus obtains a natural solution
to the hierarchy problem in this model: exponential generation of
large/small numbers explains the weakness of the observed gravitational
interactions. Of course, this is achieved at the cost of ($a$) tuning
brane tensions with the bulk cosmological constant and ($b$) assuming
negative brane-tension. It is fair to say that the RS model successfully
fuses the gauge hierarchy problem with the cosmological constant problem,
and presumably both have a common solution.

Phenomenologically, the RS model is rather similar to the ADD model, but
there are two important differences. These are
\begin{itemize}
\item Each Kaluza-Klein excitation of the bulk graviton has a 
mass~\cite{Wise,Davoudiasl}
\begin{equation}
M_n = x_n {\cal K} e^{-{\cal K}R_c\pi} \equiv x_n m_0
\end{equation} 
where $m_0 = {\cal K} e^{-{\cal K}R_c\pi} \sim 100$~GeV is the graviton
mass scale and $x_n$ are the zeros of the Bessel function $J_1(x)$ of
order unity ($n \in \mathbf{Z}$). This means that the Kaluza-Klein
gravitons have masses of a few hundred GeV, unlike the ADD case, where the
masses start from $\sim 1~\mu$eV.
\item Each Kaluza-Klein excitation of the bulk graviton couples to matter
as~\cite{Davoudiasl}
\begin{equation}
\kappa e^{{\cal K}R_c\pi} = \frac{4\sqrt{\pi}}{M_P} e^{{\cal K}R_c\pi}
\equiv \frac{4\sqrt{\pi}c_0}{m_0}
\end{equation}
where $\kappa = \sqrt{16\pi G_N}$ and $c_0 = {\cal K}/M_P \simeq 0.01 -
0.1$ is an effective coupling constant, whose magnitude is fixed by ($a$)
naturalness and ($b$) requiring the curvature of the fifth dimension to be
small enough to consider linearized gravity on the `visible' brane.
\end{itemize}

RS gravitons, thus, resemble weakly-interacting massive particles (WIMPs)  
in most models, except for ($a$) the fact that there always exists a tower
of graviton Kaluza-Klein modes and ($b$) these are spin-2 particles. In
phenomenological studies of the RS model, the mass scale $m_0$ and the
ratio $c_0$ may be treated as free parameters\footnote{The alternative
choice of $\Lambda_\pi = \overline{M}_P ~e^{-{\cal K}R_c\pi} =
m_0/\sqrt{8\pi} c_0$ instead of $m_0$ and of ${\cal K}/\overline{M}_P =
\sqrt{8\pi} c_0$ instead of $c_0$ may also be found in the
literature~\cite{Davoudiasl}.}: they are convenient replacements for the
fundamental quantities ${\cal K}$ and $R_c$:
\begin{equation}
{\cal K} = c_0 M_P \ , \qquad R_c = 
\frac{1}{\pi {\cal K}} \log \frac{{\cal K}}{m_0}
\end{equation}

Accelerator searches for RS gravitons mainly depend on the resonant
structure of graviton-mediated cross-sections. The non-observation of any
such effects at LEP-2, for example, constrains $M_1 > 206$~GeV, or $m_0 >
54$~GeV. At a hadron collider, such as the Tevatron or the LHC, one
looks~\cite{Davoudiasl} for $s$-channel graviton exchange contributions to
simple processes like $p + p(\bar p) \to \ell^+\ell^-$ or $p + p(\bar p)
\to \gamma\gamma$ or the more complicated $p + p(\bar p) \to$~dijets,
where one should expect to see graviton resonances in the invariant mass
distribution of the final state. The observation of such resonances would
certainly be a signal for the RS model. However, it may not be possible to
clinch the issue of whether such signals are uniquely due to RS gravitons
or to some other form of new physics. The reason is simple: {\it any}
particle exchanged in the $s$-channel for the above processes, such as,
for example, a $Z'$-boson, would induce very similar signals. In order to
verify that an observed resonance is indeed a graviton resonance, we need
to ($a$) measure the position, width and height of the resonance peak, and
($b$) compare the angular distribution of the final states with that
predicted for a spin-2 exchange in the $s$-channel. These may, indeed, be
possible\cite{Davoudiasl}, despite the presence of large irreducible
backgrounds at a hadron collider, which induces large errors in the width
and height measurements. It is also possible that we would see signals for
spin-2 exchanges without observing a clear resonance structure, in which 
case, one
could infer either an ADD-type scenario with closely-spaced graviton states, 
or a RS-type scenario with smeared-out resonances.
To clinch the issue, therefore, it seems natural to
turn to a high-energy $e^+e^-$ collider, where the environment is clean
and to see if one can obtain an unambiguous signal for RS gravity at such
a machine.

Several studies of RS gravitons at $e^+e^-$ colliders may already be found
in the literature~\cite{Rizzo,Das,Ghosh}. For example, one can have
$s$-channel graviton exchanges in $e^+e^- \to \mu^+\mu^-, e^+e^-$ and
$\gamma\gamma$. In these studies~\cite{Hewett} it is shown that if the
center-of-mass energy is run over a typical range, say 350~GeV to 1.5~TeV,
one gets beautiful resonances at the masses of the RS gravitons, for a
suitable choice of parameters. Conversely, if one fixes the energy and
runs over the graviton mass parameter $m_0$, one again generates
resonances whenever the condition $\sqrt{s} \simeq M_n$ is satisfied.
However, one must remember that if the RS model is true, then Nature has
only one value of $m_0$. It is also likely that the next generation
$e^+e^-$ machine will be run (like the LEP) only at certain fixed values
of beam energy~\cite{Tesla}. It is quite possible, therefore, that the
actual machine energy will be an {\it off-resonance point} for the RS
model. In such a case, the only way to excite resonances would be through
`radiative return' type reduction of the effective centre-of-mass energy
as a consequence of initial-state radiation (ISR) or beamstrahlung. As
this would automatically lead to a certain degree of suppression of the
signal (the farther away from the resonance point, the lower the flux and 
hence the greater the suppression), it may not, then, be easy to distinguish 
graviton contributions
from those due to other forms of `new physics', such as supersymmetry or
extra gauge bosons~\cite{Rizzo2}. Even if such signals are observed, it
would be useful to have a complementary signal from some other process,
which could substantiate any claim to have seen graviton exchange.

In this article, we examine the $2 \to 3$-body process $e^+ e^- \to
\gamma\nu\bar\nu$, which is observable as a final state with a single hard
transverse photon with unbalanced (missing) energy. Such states are very
distinctive and are, in fact, predicted in a wide variety of models beyond
the SM~\cite{Pandita,Datta,Ghoshal}. It is likely that some part of the
experimental effort at a high-energy $e^+ e^-$ collider will be devoted to
an analysis of this signal, which is simple, clean and physically
interesting.  The process can be thought of as a neutrino pair-production
with an ISR photon, which is, however, tagged. Since this photon carries
away a variable amount of energy, it is possible for the remaining system
to strike a $s$-channel graviton resonance, just as ISR at LEP-2 has been
seen to cause a `radiative return' to the $Z$-boson pole. The resonant
gravitons can be thought of as real particles, in which case, the basic
process is the $2 \to 2$-body process $e^+ e^- \to \gamma G_n$.  It
follows that the photon energy will be uniquely fixed by the well-known
formula 
\begin{equation} 
E_\gamma = \frac{s - M_n^2}{2\sqrt{s}}
\label{resonance} 
\end{equation} 
The photon spectrum may be expected to show Breit-Wigner resonance peaks 
at positions given by the above equation for $n = 1,2,\dots$ Observation 
of such resonances could constitute a clear signal of RS gravity.

In practice, however, the graviton resonances need not be very narrow.  
As the RS graviton decay widths vary as $c_0^2$ (see
Equation~\ref{width}), in the limit of large $c_0$, the decay widths are
very large and the resonance peaks get smoothed out till they are no
longer identifiable as resonances. One only sees a continuous photon
spectrum (except for the $Z$-boson line, of course) with a clear excess
over the SM. This is also precisely the kind of signal one would expect
for the ADD model through the process $e^+ e^- \to \gamma G_n$, where the
ADD gravitons escape detection~\cite{Giudice,Peskin}.  In fact, as we
shall see, the angular distribution of the photon also looks identical in
both these models, and it seems difficult to disentangle the signals. We
show, however, that by combining the signal for $e^+ e^- \to \gamma
\not{\!\!E}$ with that for the `benchmark' process $e^+ e^- \to \mu^+
\mu^-$, we can still obtain a clear separation between the two models.
This is an issue which has not been addressed before in the literature.

The fact that graviton widths can be large also means that we cannot, in
general, evaluate the cross-section for $e^+ e^- \to \gamma \nu \bar{\nu}$ 
using the narrow-width approximation for graviton propagators. The full 
machinery of a $2 \to 3$-body process must be used.

In the following section we briefly discuss the decays of RS gravitons and
then go on to describe the long and cumbersome calculation of the $2 \to
3$-body process $e^+ e^- \to \gamma \nu \bar \nu$. Our numerical results,
including discovery limits in the RS model parameter space, are presented
in Section 3. In Section 4, we discuss how to disentangle signals for the
RS model in the large $c_0$ limit from those for the ADD scenario.  We
summarize our results in Section 5 and relegate some important formulae to
the Appendix.

\section{Graviton Resonances and the Process $\bf{e}^{\bf +} {\bf e}^{\bf -} 
\mathbf{\to \gamma \nu \bar \nu}$}

As explained above, experimental searches for RS gravitons mainly focus on
the fact that they can form narrow resonances in high-energy scattering
processes. A simple calculation of the graviton width yields the form
\begin{equation}
\Gamma_n = c_0^2 x_n^3 m_0 \sum_P \Delta^{(n)}_{P\bar P}
\label{width}
\end{equation}
where the sum $\sum_P$ runs over all pairs of particles ($G_n \to P\bar
P$)  and the $\Delta^{(n)}_{P\bar P}$ are dimensionless functions of $x_n$
and the ratios $r_P = m_P/m_0$. Their exact forms are listed in the Appendix.

In Figure~1, we show the region in the $c_0$--$m_0$ plane for which the
graviton resonances have widths less than 10\% of the mass. This 
is a reasonable estimate to identify a {\it narrow} resonance, taking likely
experimental resolutions into account~\cite{Tesla}.  It is
clear from the figure that, for low values of $c_0$, the graviton widths
remain quite small, leading to identifiable resonances. This is true for
$n = 1$ for a large range of the parameter space considered, unless
$c_0 > 0.06$. It is likely, therefore, that the first massive graviton 
state will exhibit an identifiable resonance peak, a fact which
forms the basis of most existing studies in the context of the
LHC~\cite{LHC}.

The situation for the next two resonances is not so simple. The
figure makes it clear that we can expect fairly sharp resonance peaks
for $n = 2, 3$ for values of $c_0$ below 0.03 and 0.02 respectively.
However, as $c_0$ increases, these widths grow steeply and soon the 
resonance shape is lost. For $n > 3$ this seems to be the case, even 
when $c_0$ is small. We thus see that for suitable values of the parameters, 
it might be just possible to observe the first three graviton resonances 
in the RS model, but hardly more. In fact, most of the time, only two 
 --- or even one --- resonance(s) will be identifiable. Moreover, any 
analysis based on identifiable resonances, such as those in Ref.~\cite{LHC} 
will break down when all the resonances are wide, e.g. for $c_0 > 0.06$.

As the exact conditions for identifying a resonance at a linear collider 
would depend crucially on
experimental resolutions, it is premature to set up a definite criterion for 
resonance identification. Ideally, we should use something like the 
$Z$-lineshape analysis at LEP-1, and presumably some such study will be done 
if and when we have data. For the present, the condition 
$\Gamma_n < \frac{1}{10}M_n$ was chosen purely for illustrative purposes and 
the following discussion is {\it not} predicated on any such numerical 
criteria. 

\begin{figure}[htb]
\begin{center}
\vspace*{3.6in}
      \relax\noindent\hskip -4.8in\relax{\includegraphics{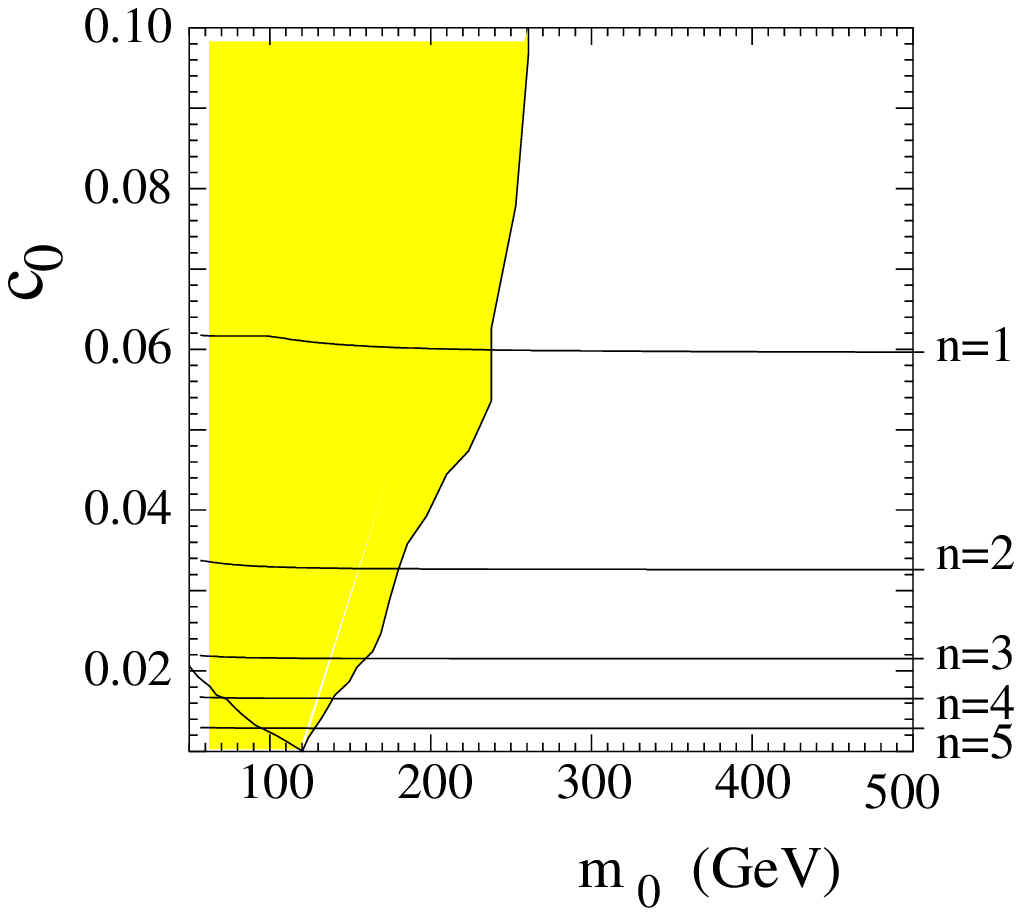}}
\end{center}
\end{figure}
\vspace*{-0.5in}
\noindent {\bf Figure~1}.
{\footnotesize\it Illustrating the region (below the solid lines) in the
parameter space of the RS model for which the graviton width remains small
($\Gamma_n < \frac{1}{10}M_n$), for the first five graviton resonances $n
= 1, 2, 3, 4, 5$. The shaded region represents the constraints from
dilepton production at the Tevatron Run I.}
\vskip 10pt

Along with the graviton widths, in Figure~1 we have also shown the results
of a phenomenological analysis~\cite{Hewett} of Tevatron dilepton data
constraining the RS model parameters. The small negative-slope line inside
the shaded region indicates constraints from precision data, but are not
very important since they are subsumed in the region excluded by the 
dilepton data. We note that, for $c_0 \geq 0.01$, we already have 
$m_0 \geq 125$~GeV, corresponding to setting all graviton resonances 
heavier than about 480~GeV. This means that RS gravitons are practically 
inaccessible for a 500~GeV linear collider and one has to consider higher 
energies to excite graviton resonances.

We now concentrate on the $2 \to 3$-body process
$$
e^-(k_1, \lambda_1) + e^+(k_2, \lambda_2) 
\longrightarrow 
\gamma(p_1) + \bar{\nu}(p_2) + \nu(p_3)
$$
which is observable as a final state with a single hard photon with
unbalanced (missing) energy. Note that we display only the helicities of
the initial states, as the (unobserved) helicities of the final states
will be summed over. 

\newpage
\begin{figure}[htb]
\begin{center}
\vspace*{6.8in}
      \relax\noindent\hskip -5.8in\relax{\includegraphics{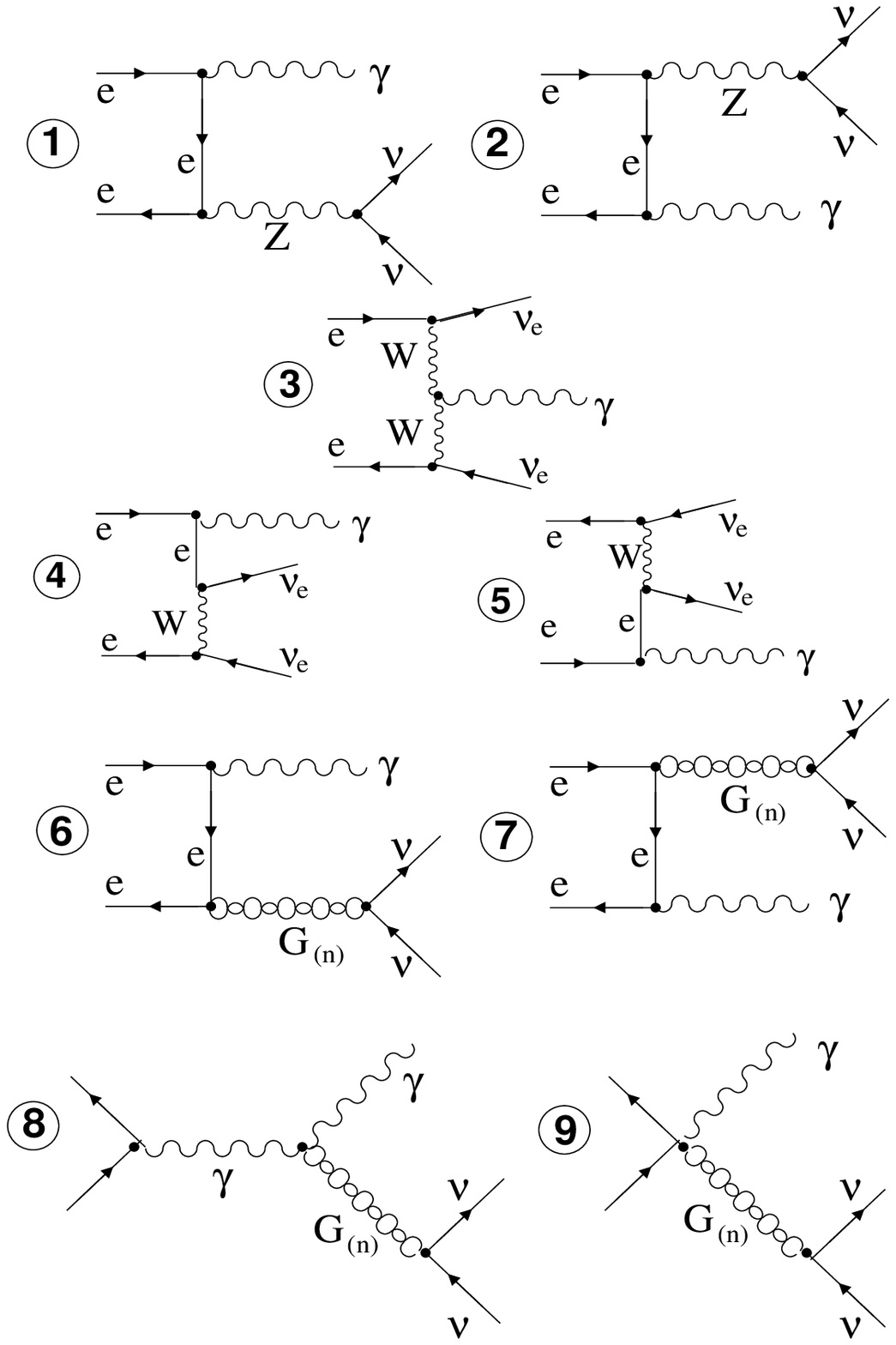}}
\end{center}
\end{figure}
\vspace*{-0.3in}
\noindent {\bf Figure~2}.
{\footnotesize\it Feynman diagrams contributing to the process $e^+e^- \to
\gamma \nu\bar{\nu}$ in the RS model. Diagrams numbered $1$ to $5$ are
present in the SM model, while diagrams $6$ to $9$ involve graviton
exchange. Diagrams numbered $3$, $4$ and $5$ exist only for the electron
neutrino.}
\vskip 10pt

Figure~2 shows the complete set of tree-level Feynman
diagrams contributing to the process in the RS model. 
Diagrams 1--2, 6--9
are labeled with $\nu$ indicating a neutrino of {\it any} flavour, while
diagrams 3--5 are specifically labeled with the electronic flavour
$\nu_e$. Feynman amplitudes corresponding to these nine diagrams are given
in the Appendix.

We use these amplitudes to calculate the cross-section in terms of the
usual $2 \to 3$~body invariants $s, s_1, s_2, t_1$ and $t_2$. The
expressions, including 36 possible interference terms, are long and messy.
In the interests of brevity, we do not present them in this paper. The
final cross-section was coded into a Monte Carlo event generator, which
produced the numerical results.  The graviton contributions require a
summation over graviton Kaluza-Klein modes, parametrised by the function
$\Lambda(Q^2)$, which is defined as
\begin{equation}
\Lambda(Q^2) = \sum_n \frac{\kappa^2 e^{2\pi{\cal K}R_c}}
                           {Q^2 - M_n^2 + i M_n \Gamma_n}
\end{equation}
where $n$ runs over all the graviton modes. We now explain briefly how the
function $\Lambda(Q^2)$ is evaluated.

As mentioned above, masses of the Kaluza-Klein gravitons in the RS model
vary as $M_n = x_n m_0$ where $x_n$ are the zeroes of the Bessel function
$J_1(x)$. These may be approximated by~\cite{Abramowitz}
\begin{equation}
x_1 \simeq 1.22\pi \ ,  \qquad\qquad  x_2 \simeq 2.23\pi  \ , \qquad\qquad
x_n \approx x_2 + (n - 2) \pi \ .  
\end{equation}
However, we have seen that the width $\Gamma_n$ is a complicated function
of the mass $M_n$. If we consider Breit-Wigner resonances up to $n = N$
and simple propagator summation thereafter, we get a simplified form
\begin{eqnarray}
\Lambda(Q^2) & \approx & 16\pi \frac{c_0^2}{m_0^2} 
\left[ \sum_{n=1}^N \frac{1}{Q^2 - M_n^2 + i M_n \Gamma_n}
+\sum_{n=N+1}^\infty \frac{1}{Q^2 - M_n^2}   \right]
\\
& = & 16\pi \frac{c_0^2}{m_0^4} 
\left[ \sum_{n=1}^N \frac{1}{x^2 - x_n^2 + i x_n \gamma_n}
+ \psi\left(\frac{x_2+x}{\pi} + N - 1\right) 
- \psi\left(\frac{x_2-x}{\pi} + N - 1\right) \right]
\nonumber
\end{eqnarray} 
where $x^2 = Q^2/m_0^2$, $\gamma_n = \Gamma_n/m_0$ and $\psi$ is the Euler
digamma function, which is obtained on performing the infinite sum of the
remaining terms\footnote{Strictly speaking, the sum is not infinite, but
must be cut off at the five-dimensional Planck scale. However, this is so
much larger than the experimental energy that the sum is, for all
practical purposes, infinite.}. In our numerical analysis, we determine
$N$ by the simple criterion that for $n > N$, the Breit-Wigner
contribution from a single resonance should not exceed 1\% of the
resonance value of $\Lambda(Q^2)$. Calculated in this way, we have checked
that our results closely match those of, for example, Ref.~\cite{Hewett}.

\section{Results and Discussion}

Our numerical analysis of the problem has been performed for two values of
center-of-mass energy, viz., $\sqrt{s} = 1$~TeV and $\sqrt{s} = 2$~TeV. As
explained above, Tevatron data already rule out graviton resonances 
accessible at a 
500~GeV collider, and hence we have carried out our analysis for higher 
energies, such as are contemplated for a variety of linear collider designs. 
Noting that the final state consists of a single hard isolated 
photon, we impose the following kinematic cuts

\begin{itemize} 
\item The photon should have energy $E_\gamma \geq 20$~GeV. 
\item The photon scattering angle $\theta_\gamma$ should satisfy $15^0
\leq \theta_\gamma \leq 165^0$. 
\end{itemize} 

These ensure that the tagged photon does not arise from beamstrahlung or
other similar sources~\cite{Drees}.  Assuming 100\% efficiency in photon
detection (we shall see later that the actual efficiency factors cancel
out), we then calculate the differential and total cross-section for both
($a$)  unpolarised and ($b$) polarised beams.

\begin{figure}[htb]
\begin{center}
\vspace*{3.2in}
      \relax\noindent\hskip -5.8in\relax{\includegraphics{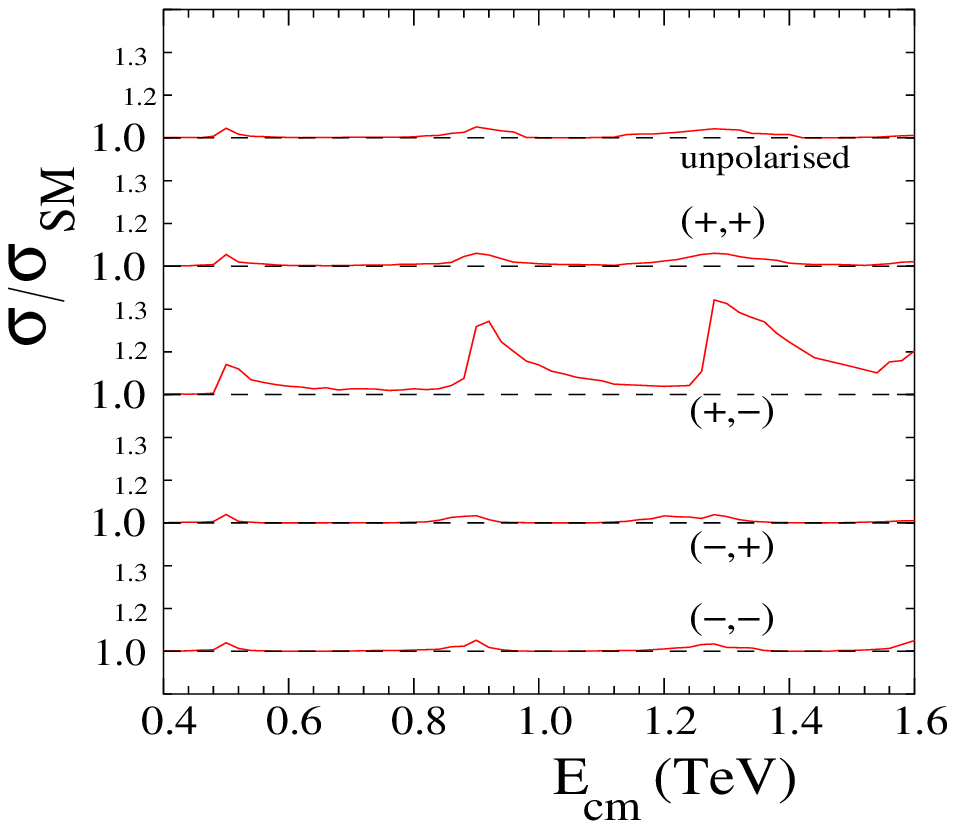}}
\end{center}
\end{figure}
\vspace*{-0.4in}
\noindent {\bf Figure~3}.
{\footnotesize\it Illustrating the variation of the ratio between signal
$\sigma$ and background $\sigma_{SM}$ with center-of-mass energy
$\sqrt{s}$ for different choices of {\rm [sgn(P$_e$), sgn($P_p$)]}.  
Clearly the best choice is to take ${\cal P}_e = 0.8$ and ${\cal P}_p =
-0.6$. The signal was calculated taking $m_0 = 125$~GeV and $c_0 = 0.01$,
which is just allowed by the Tevatron constraints shown in Figure~1. }
\vskip 10pt

Beam polarisation can be an extremely efficient tool in reduction of the
SM background (diagrams 1--5 in Figure~2). This is because the largest
contributions to the signal arise from $t$-channel exchange of $W$-bosons
in the process $e^+e^- \to \gamma \nu_e \bar{\nu}_e$ (diagrams 3--5).
These diagrams are strongly suppressed if the electron (positron) beam is
right (left) polarised, because of the $V - A$ nature of the $W$-boson
coupling. This is beautifully illustrated in Figure~3, where we plot the
total cross-section for $e^+e^- \to \gamma \nu \bar{\nu}$ 
(signal-to-background ratio $\sigma/\sigma^{SM}$)
against the center-of-mass energy $E_{cm} = \sqrt{s}$. 
Figure~3 actually comprises five graphs, viz., 
the unpolarised (0,0) case, as well as all four choices of
[$sgn(P_e), sgn(P_p)$] taking typical values~\cite{Tesla} $|P_e| = 0.8$
and $|P_p| = 0.6$. The choice (+,-), corresponding to
right (left) polarised electron (positron), produces roughly an
order-of-magnitude suppression of the background and thereby throws the
new physics signal into prominence in a way that cannot be achieved
otherwise. For the rest of this paper, then, we concentrate on this
choice. It turns out that even with the strong background suppression, 
the total cross-section is still in the range of a few hundred
femtobarns. As we expect~\cite{Tesla} a luminosity of at least
10~fb$^{-1}$ at a linear collider, this would mean a few thousand events, 
and more if the high-luminosity option of 1000~fb$^{-1}$ is achieved. 
Thus, the use of beam polarisation turns out to be an undiluted blessing.
 
\begin{figure}[htb]
\begin{center}
\vspace*{6.5in}
      \relax\noindent\hskip -5.8in\relax{\includegraphics{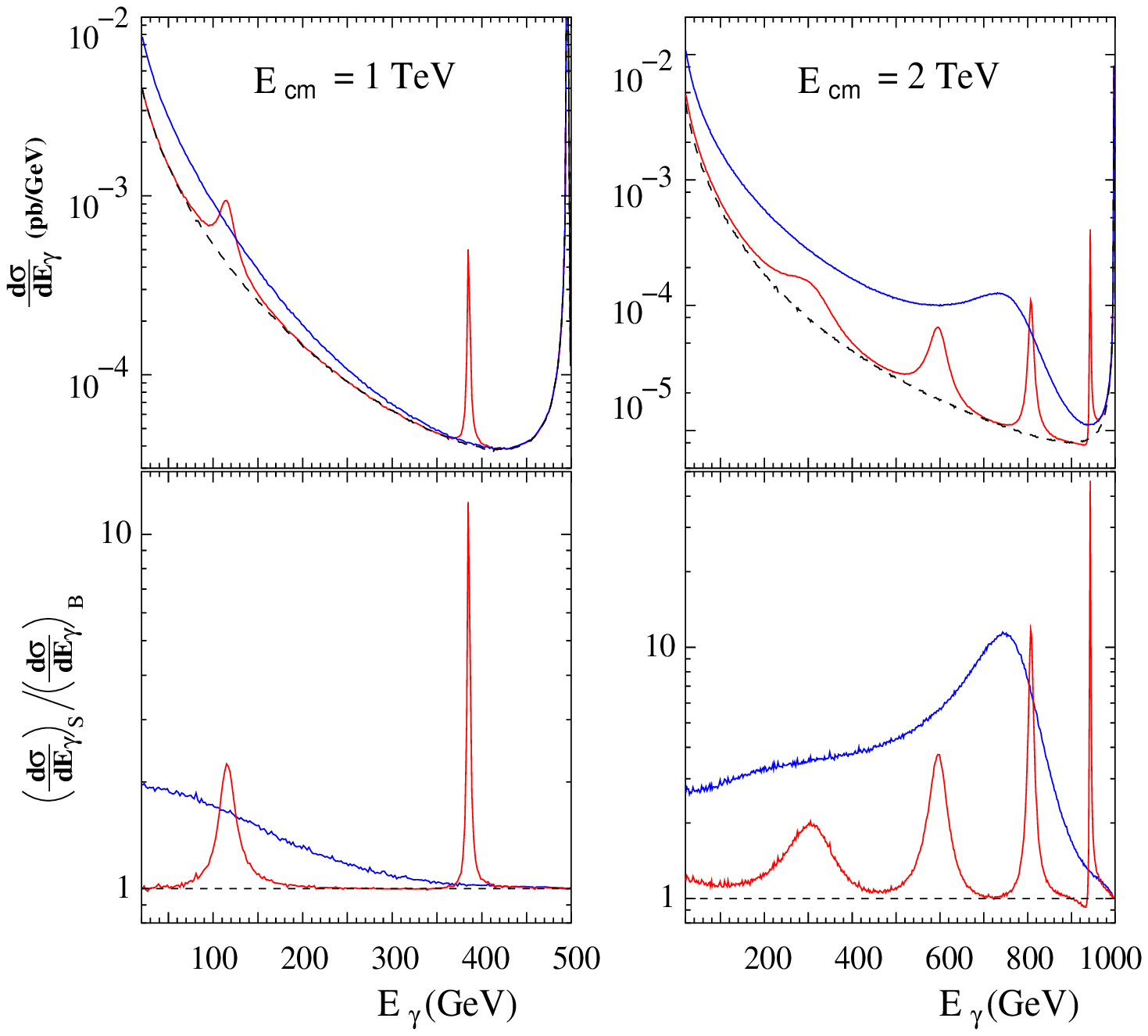}}
\end{center}
\end{figure}
\vspace*{-1.5in}
\noindent {\bf Figure~4}.
{\footnotesize\it Energy spectrum of the tagged photon for the 
Tevatron-allowed parameter
choices $m_0 = 125$~GeV, $c_0 = 0.01$ (red)  --- corresponding to narrow
resonance(s) --- and $m_0 = 250$~GeV, $c_0 = 0.07$(blue) --- corresponding
to broad, indistinct resonances. In the ordinate labels, {\rm S} and
{\rm B} denote signal (SM plus gravitons) and background (SM only)  
respectively. Note that the right-most peak (almost flush with the edge 
of the box) in the upper graphs, which is due to the
$Z$-boson, can be removed by taking the {\rm S/B} ratio. 
We consider polarised beams with ${\cal P}_e = 0.8$ and ${\cal P}_p = -0.6$. }

We can study the kinematic distributions of the only two measurables: the
energy $E_\gamma$ and the scattering angle $\theta_\gamma$ of the photon.
In Figure~4, we display, for $\sqrt{s} = 1$ and 2~TeV respectively,
the photon energy spectrum for both the signal and the SM background. The
dashed (black) line represents the background (note the $Z$-boson peak
at the extreme right), while solid curves correspond to the signal for 
low (red) and high (blue) values of the parameter $c_0$. Sharp resonances
are obtained with the parameter choice $c_0 = 0.01$ and $m_0 = 125$~GeV,
which is the lightest mass spectrum allowed by the Tevatron constraints. 
It corresponds to graviton resonances with $M_n \simeq 479, 877, 1272$
and 1665~GeV for $n = 1,2,3$ and 4 respectively. Only the first two are
kinematically accessible at a 1~TeV machine, but all four will be accessible
if the centre-of-mass energy rises to 2~TeV. Observe
that the resonance peaks broaden as the order $n = 1,2, \dots$ of the
Kaluza-Klein excitation increases. For a large values of $c_0$,
viz. $c_0 = 0.07$ (with $m_0 = 250$~GeV, to be consistent with the Tevatron
constraints -- this doubles the graviton mass), 
the resonance line-shape merges into a continuous spectrum.

In the upper halves of the two graphs in Figure~4, we display the
differential cross-section for the process $e^+e^- \to \gamma + \not{\!\!
E}$. The bottom halves show the same distribution, except that now we
exhibit the signal-to-background (S/B) ratio. Not only does this remove
the uninteresting $Z$-peak, but it also takes care of any radiative
corrections, efficiency factors, etc, which can be written in a
factorisable form.

In Figure~5, we exhibit similar signal-to-background ratios for larger
values of $m_0$, when the resonances become heavier. The first three
graviton resonances lie at 766~(958), 1403~(1754), 2035~(2543)~GeV
respectively for $m_0 = 200~(250)$~GeV. This means that only the first
two resonances would be kinematically accessible, even when the
centre-of-mass energy is 2~TeV. While we get sharp peaks (red curves) 
for lower values of $c_0$, when we shift $c_0$ close to the maximum value 
allowed by Tevatron data, the peaks get smeared out, except for the
first one, which still retains a more-or-less recognisable shape.  

\bigskip

Thus, we can expect one of the following possibilities.

\begin{itemize}
\item Case I:  {\it Two (or more) clear resonances are seen in the photon
spectrum,} or rather, in the signal-to-background ratio. This
would correspond to a relatively low value of $m_0$ and a small value of
$c_0$, and would constitute a strong hint of RS gravity. For confirmation,
we need to check two more details. First, the peak at smaller $E_\gamma$
should be lower and broader than that at larger $E_\gamma$, as illustrated in
Figures~4 and 5. This is because $\Gamma_n \propto x_n^3$. Secondly, the
positions of the two resonances will bear a definite relation if they are
due to RS gravitons. Using Equation~(\ref{resonance}) this works out to
the requirement that, for resonance values $E_\gamma^{(1)} >
E_\gamma^{(2)}$, 
\begin{equation}
\sqrt{\frac{\sqrt{s} - 2E_\gamma^{(1)}}{\sqrt{s} - 2E_\gamma^{(2)}}}
= \frac{M_1}{M_2} = \frac{x_1}{x_2} \simeq 0.546 \ .
\end{equation}
It would be a very remarkable coincidence,
indeed, if some other form of new physics\footnote{Barring variations of
the RS model which put SM fields in the bulk~\cite{universal}. 
The authors are grateful to
K.~Sridhar for pointing out this feature.} --- such as, for example, two
extra $Z'$ bosons --- could reproduce this ratio. Thus, if we do see two
clear resonances, {\it satisfying the above relation}, it may be quite in 
order to claim this as a `smoking gun' signal for the RS model (or a close
variation).

\begin{figure}[htb]
\vspace*{4.8in}
\begin{center}
      \relax\noindent\hskip -5.8in\relax{\includegraphics{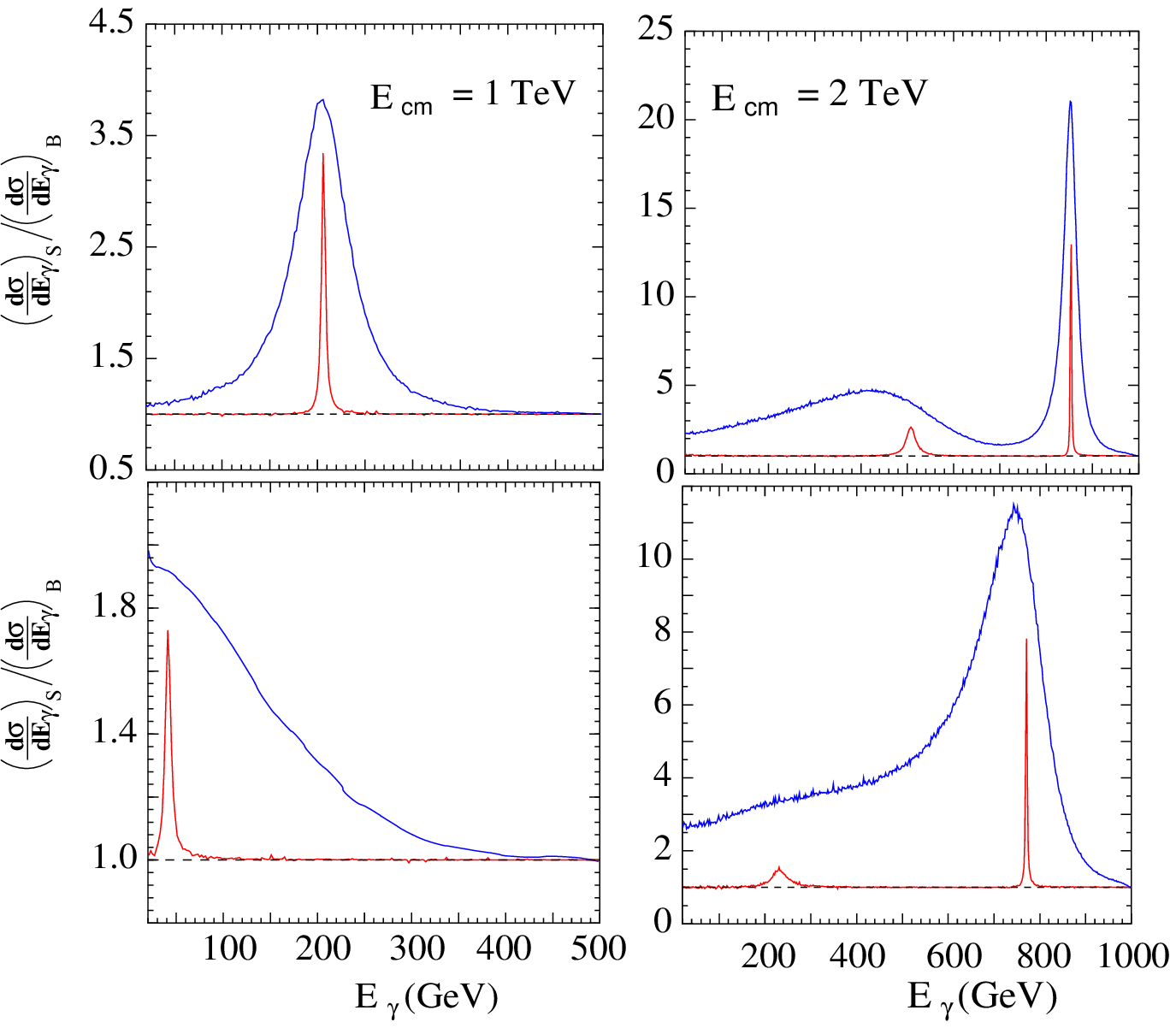}}
\end{center}
\end{figure}
\vspace*{-0.6in}
\noindent {\bf Figure~5}.
{\footnotesize\it Illustrating the shift and disappearance of resonance
peaks with increasing $m_0$ and $c_0$. The upper graphs correspond to $m_0
= 200$~GeV, $c_0 = 0.01$~(red) and $c_0 = 0.04$~(blue) while the lower
ones correspond to $m_0 = 250$~GeV, $c_0 = 0.01$~(red) and $c_0 =
0.07$~(blue). Notations and conventions are similar to those used in 
Figure~4. All parameter points are allowed by Tevatron data.}
\vskip 10pt

\item Case II: {\it One sharp resonance is seen, but no more.} If this is
a graviton resonance, it will correspond to a somewhat larger value of
$m_0$ when only the first (lightest) graviton resonance is kinematically
accessible. A single resonance could also indicate the presence of some
other kind of new physics, such as, for example, an extra $Z'$ boson, or
perhaps a leptophilic scalar. Of course, one can still measure the
position, width and height of the resonance, which would be fitted with
just two parameters $m_0$ and $c_0$. A consistent fit would constitute
circumstantial evidence for RS gravitons, but would hardly be conclusive.
However, in the case of RS gravitons, we should also expect extra
contributions to 2-body final states such as $\mu^+\mu^-$, and the angular
distribution of these might be useful to distinguish between different
types of new physics. It is also worth mentioning that such measurements
at a linear collider would be complementary to earlier measurements (including
that of the spin) at the LHC and the combined evidence could be quite 
conclusive.

\item Case III:  {\it No resonances are seen, but large excesses in the
S/B ratio are observed.} This could be due to RS gravitons with a large
coupling $c_0$, which means smeared-out resonances, and a mass scale such
that even the first resonance is 
inaccessible to the linear collider. However, it could also
be due to various other sources, such as ADD gravitons (which form a
near-continuous spectrum), or, perhaps the production of supersymmetric
particles. A process like $e^+ e^- \to \gamma \widetilde{\chi}_1^0
\widetilde{\chi}_1^0$, where $\widetilde{\chi}_1^0$ is the (invisible)
lightest neutralino, would contribute~\cite{Pandita,Datta} an excess very
similar to the observed one. In this case, it may be a real challenge to
determine the nature of the new physics causing this signal. In the next
section we take up the issue of separating ADD versus RS graviton signals
with comparable kinematic signatures. However, the comparison with
supersymmetry and other types of new physics is a tricky issue, very much
dependent on the choice of model parameters, and it is premature, at this
point of time, to take up a detailed investigation.

\item Case IV: {\it No excess is seen: the observed spectrum is completely
consistent with the SM prediction.} This possibility would be rather
disappointing, but it can hardly be ignored for that reason. In such a
case, we would argue that the RS gravitons are far too heavy to be
kinematically accessible to the linear collider. We would then get as
bounds what we now present in Figure~6 as 95\% confidence level discovery
limits. 
\end{itemize}

In order to obtain the discovery limits shown in Figure~6, we have to take
into account any excess in the differential cross-section ratio shown in
Figures~4 and 5, {\it irrespective of whether a resonance is discernible or
not}. To do this, we make a simple-minded $\chi^2$ analysis of the bin-wise
energy distribution of the photon, and indicate the region of the
parameter space where a departure from the SM prediction would be
observable at 95\% confidence level. We define a $\chi^2$ by 
\begin{equation}
\chi^2(c_0,m_0) = \sum_i \frac{[N_i(c_0,m_0) - N_i^{SM}]^2}{N_i^{SM}}
\end{equation}
assuming Gaussian statistical errors\footnote{At this stage, we consider
it reasonable to ignore possible systematic errors.} and defining 
$N_i = {\cal L} \sigma_i$ as the 
number of events expected in the $i$th bin. We then require $\chi^2$
to be greater than the number expected for random Gaussian fluctuations
of the SM expectation. Figure~6 clearly illustrates how far
a high energy linear collider can probe the RS model. Compared with the
bound from Tevatron Run-I dilepton data~\cite{Davoudiasl,Hewett} we see
that a linear collider can probe much larger values of $m_0$, i.e.
graviton resonances which are many times heavier. It is worth pointing
out that the mass range which can be probed at a 2~TeV machine is 
significantly greater than that which can be accessed at a 1~TeV machine.

\begin{figure}[htb]
\begin{center}
\vspace*{4.0in}
      \relax\noindent\hskip -5.8in\relax{\includegraphics{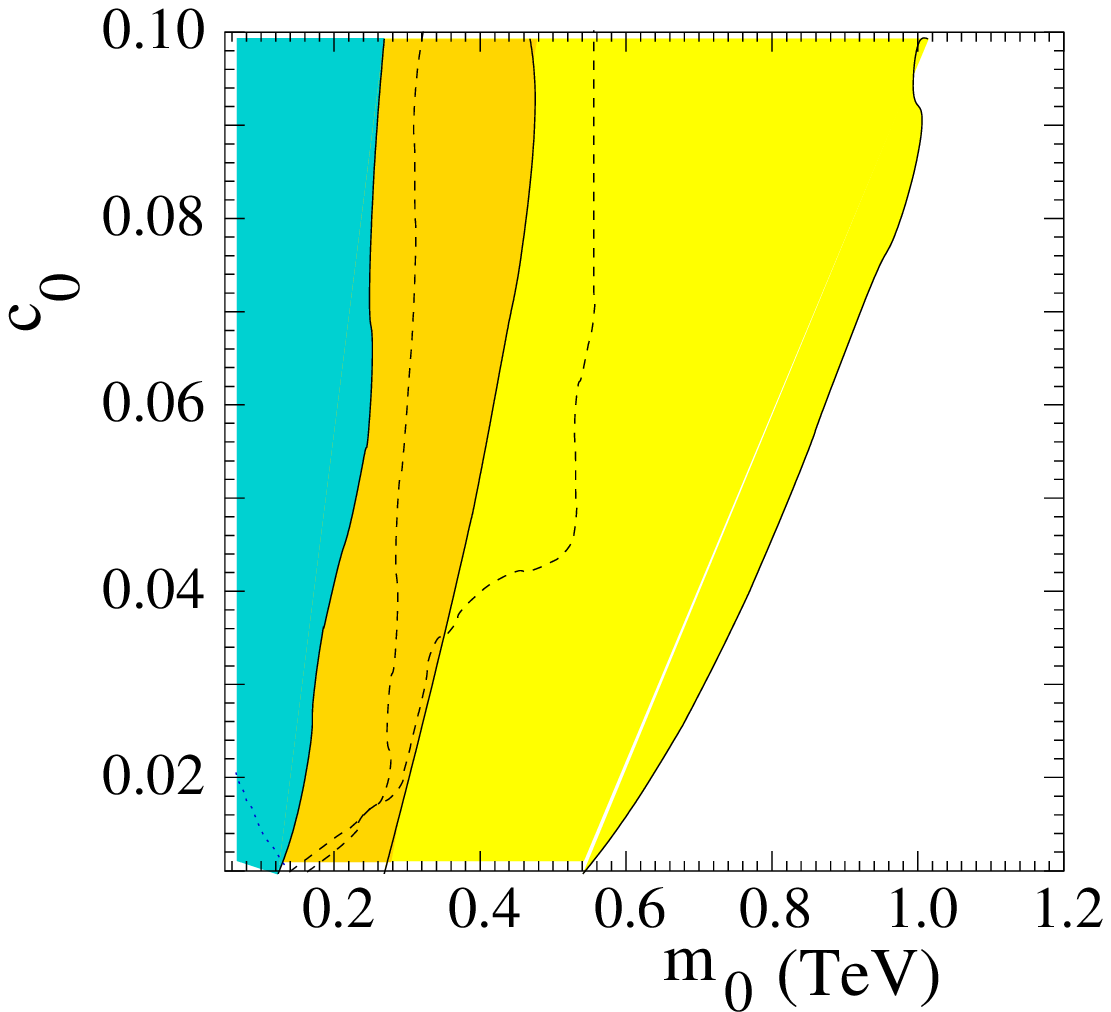}}
\end{center}
\end{figure}
\vspace*{-0.4in}
\noindent {\bf Figure~6}.
{\footnotesize\it Discovery limits at 95\% C.L. for the RS model in the
$c_0$--$m_0$ plane using single-photon signals. The region shaded
blue is the Tevatron bound from dileptons. The regions shaded light (dark)
yellow correspond to $\sqrt{s} = (1)~2$~TeV and high luminosity
(${\cal L} = 10^3$~fb$^{-1}$). Broken lines on the left~(right)
correspond to low luminosity (${\cal L} = 10$~fb$^{-1}$) and
$\sqrt{s} = 1~(2)$~TeV. }
\vskip 10pt

\section{Distinguishing between RS and ADD Models}

We now come to the crucial issue: when the mass $m_0$ and/or the coupling 
$c_0$ is large, and no
clear resonance structure is discernible, is it possible to distinguish
between the $\gamma + \not{\!\! E}$ signal arising from RS graviton from
those arising from ADD gravitons~\cite{Giudice,Peskin}? We have already
explained that not much can then be gained from a study of the photon
energy spectrum. The only other observable is the photon angular
distribution $d\sigma/d\theta_\gamma$. In the ADD model, the signal arises
from $2 \to 2$-body processes corresponding to diagrams similar to nos.
6--9 in Figure~2, but without the neutrino lines. As this involves real 
gravitons in the final state, there is no
interference with the SM diagrams and this could be expected to lead to
some difference in the angular distribution of the photon.

In Figure~7, we display ($a$) the photon energy spectrum and ($b$) the 
(normalized) photon angular distribution at a 2~TeV collider for the 
following cases:
\begin{enumerate}
\item RS gravitons, with $m_0 = 200$~GeV and $c_0 = 0.01$, indicated with a
solid red line;
\item RS gravitons, with $m_0 = 400$~GeV and $c_0 = 0.09$, indicated with a
solid blue line;
\item ADD gravitons with $d = 3$ and $M_S = 5$~TeV, indicated with a solid 
black line.
\end{enumerate} 
Considering first the energy spectrum shown in Figure~7($a$), 
case~1 shows narrow resonances and may be distinguishable from the 
other two. However, the others differ only for $E_\gamma > 400$~GeV, where
the differential cross-section is at the level of 0.1 fb/GeV. As this is 
just the region where we may expect larger errorbars, it is unlikely that 
the energy spectrum data will provide a clear distinction between cases~2 
and 3 above. Turning, therefore, to the angular distribution shown in 
Figure~7($b$), we see that the same pattern repeats itself, with a clear
demarcation between the resonant and non-resonant cases. In fact, unless
the errorbars are very small, we may not be able to distinguish between 
any of these cases at all. A more detailed study, including angular 
resolution, is required to decide this issue, but obviously cannot be 
carried out at this early stage.

\begin{figure}[htb]
\begin{center}
\vspace*{3.8in}
      \relax\noindent\hskip -5.8in\relax{\includegraphics{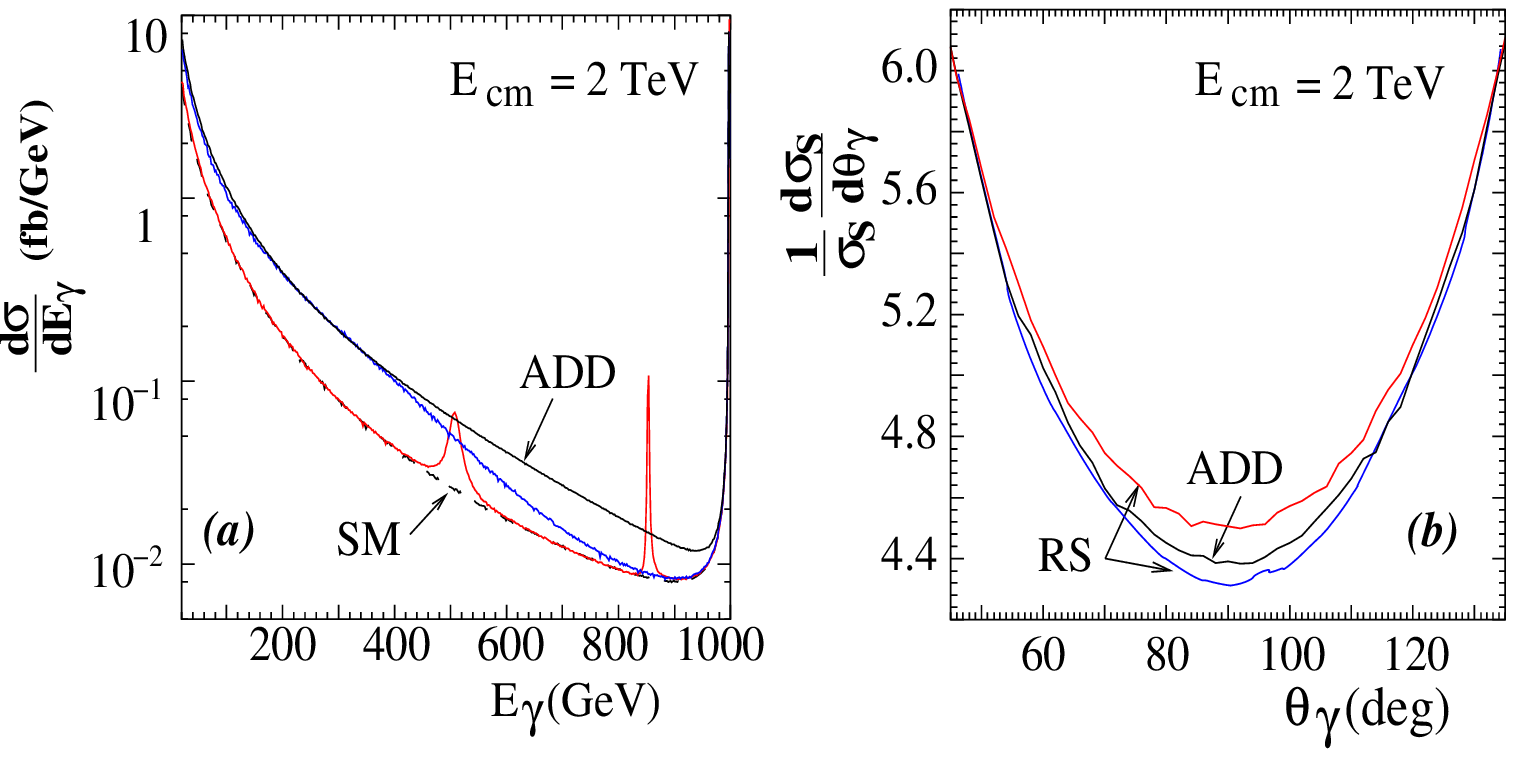}}
\end{center}
\end{figure}
\vspace*{-1.2in}
\noindent {\bf Figure~7}.
{\footnotesize\it ({\em a}) Energy spectrum of the tagged photon showing
the similarity between RS and ADD model predictions, in some regions of
the parameter space. Solid red (blue) lines correspond to parameter sets
$m_0 = 200$~GeV, $c_0 = 0.01$ ($m_0 = 400$~GeV, $c_0 = 0.09$)
respectively. Solid black lines correspond to ADD model with $d = 3$ and
$M_s = 5$~TeV. The broken line shows the SM contribution. ({\em b})
Normalized angular distribution of single photons in the RS and ADD models
for $\sqrt{s} = 2$~TeV, other conventions being the same as in ({\em a}).
The subscript {\em S} indicates that we consider the signal with the SM 
part subtracted.  }
\vskip 10pt

It appears, therefore, that a study of the process $e^+ e^- \to \gamma
\nu \bar \nu $ alone cannot clearly distinguish between the ADD and RS
models, should a continuous excess in the photon spectrum be observed.
However, if we consider this process {\it in conjunction} with a benchmark
process, like $e^+ e^- \to \mu^+ \mu^-$, for example, we do find a marked
difference. This is because the $\gamma + \not{\!\! E}$ signal arises in
the ADD model from a $2 \to 2$-body process, whereas in the RS model it
arises from a $2 \to 3$-body process, which is phase-space suppressed.
Consequently, when the cross-sections for the two match, the ADD parameter
$M_S$ must be rather large, as is evidenced by the choice 5~TeV in
Figure~7. When we consider the corresponding contributions to a $2 \to 2$
body process, like $e^+ e^- \to \mu^+ \mu^-$, the same choice of
parameters would yield a much larger cross-section for the RS model. Using
this idea as a cue, in Figure~8 we plot our predictions, at a
2~TeV machine, for $\sigma(e^+e^- \to \gamma \not{\!\! E})$ versus
$\sigma(e^+ e^- \to \mu^+ \mu^-)$ for both the models in question. 
For the RS model, we choose two values $m_0 = 200, 400$, and vary the
value of $c_0$ between 0.01 and 0.1, which is the expected range. For the
ADD model we have varied $M_S$ from $\sqrt{s}$ to much larger values, for
$d = 3, 6$. Both sets of curves converge, in the decoupling limit, to the 
SM predictions (beyond the lower left corner).

\begin{figure}[htb]
\begin{center}
\vspace*{3.9in}
      \relax\noindent\hskip -5.8in\relax{\includegraphics{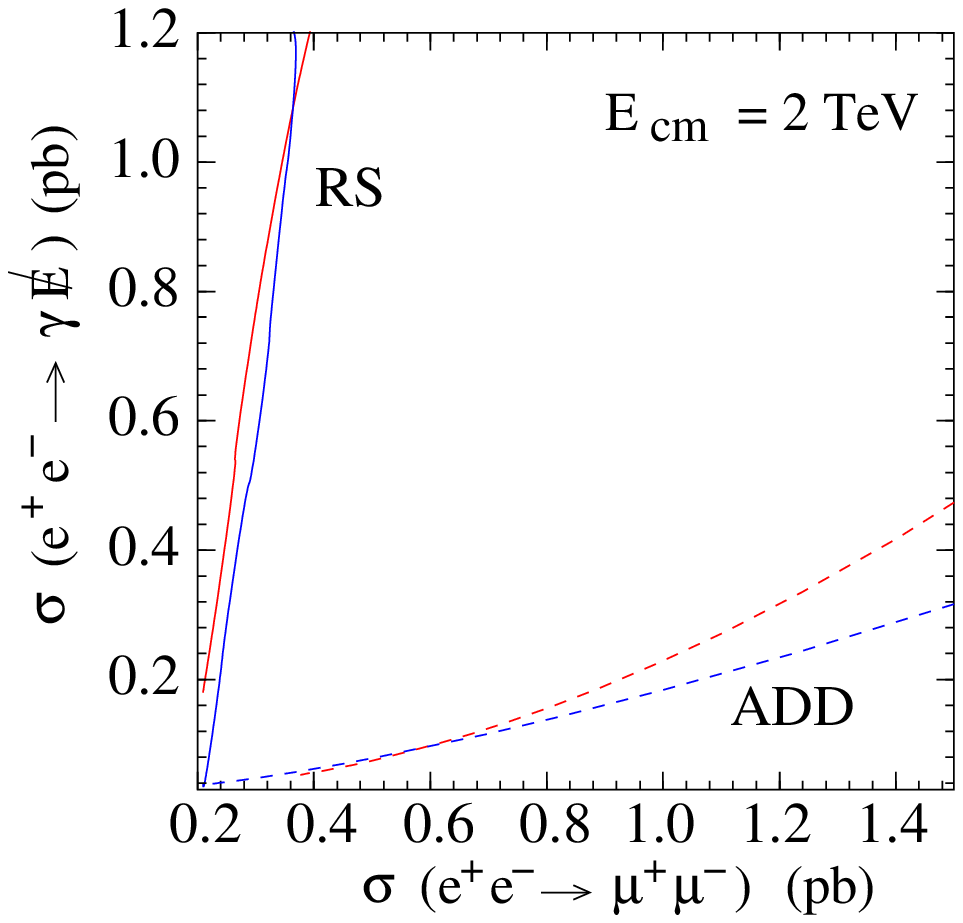}}
\end{center}
\end{figure}
\vspace*{-0.6in}
\noindent {\bf Figure~8}.
{\footnotesize\it Correlation plot showing the cross-section 
for the single photon signal vis-\'a-vis the
cross-section for muon pair-production. Broken lines correspond to the ADD
model for two values of $d = 3$~(red) and $6$~(blue), while solid lines
correspond to the RS model for two values of $m_0 = 200$ GeV~(red), $400$
GeV~(blue).}
\vskip 10pt

A glance at Figure~8 makes it obvious that a correlation plot of this
nature shows a clear difference between ADD and RS models. If the actual
observation point is found to lie toward the upper-left corner of the
graph, one could clearly say that the signal is of RS gravity. If the
observation point were to lie toward the bottom-right corner of the graph,
it would be equally certain that the signal must come from ADD gravity.
If, however, the point comes out close to the bottom-left corner, where
the two sets of graphs merge into the SM prediction, it will be difficult
to distinguish any kind of new physics effect at all. This would correspond
to the Case IV in Section III.

A correlation plot such as that shown in Figure~8 could, in fact, prove
more useful than merely in distinguishing between ADD and RS models. If we
consider a supersymmetric model, such as the minimal supersymmetric SM,
for example, we will not predict extra contributions to $e^+ e^- \to \mu^+
\mu^-$. In that case, the observed point will lie on the vertical axis of
Figure~8, and may be distinguished from the graviton contributions if the
errors are small enough. Contributions from one (or more) extra $Z'$
bosons would look similar to the RS graviton contributions, but would be
distinguishable by the existence of clear resonances in the photon
spectrum, since $Z'$ widths are usually small. We have already discussed
how to distinguish between graviton and $Z'$ resonances.

Finally, it should be noted that the process considered in this article,
namely $e^+ e^- \to \gamma \nu \bar \nu$ is not the only gravitonic process
at a linear collider. There could also be complementary 
processes~\cite{unique} such as,
for example, $e^+ e^- \to P \bar{P}(\gamma)$ where $P$ is any SM particle
that can be tagged, which would show similar resonant behaviour due to ISR 
and beamstrahlung effects. A detailed investigation of these has been taken 
up~\cite{GRR}.

\section{Summary and Conclusions}

To summarize then, it is quite evident that, like supersymmetric models,
solutions of the hierarchy problem which require the existence of extra
hidden dimensions are here to stay. It is, therefore, of great interest to
investigate their phenomenological conclusions. In this paper, we have
investigated the process $e^+ e^- \to \gamma \nu \bar \nu$ in the
(minimal) Randall-Sundrum model of warped quantum gravity. This process
has a clean signature, since the final observed state consists of a single
hard photon with unbalanced momentum. We show that the energy spectrum of
this photon could show clear resonances corresponding to massive
gravitons, and discuss how these can be distinguished from other forms of
new physics yielding resonances. We then consider the large-mass and 
large-coupling limit of the RS model, where the resonances are smeared 
out or inaccessible and the photon
spectrum is practically indistinguishable from that predicted by the ADD
model with large extra dimensions. We demonstrate that a correlation plot
between the cross-sections for $e^+ e^- \to \gamma \not{\!\! E}$ and a
benchmark process like $e^+ e^- \to \mu^+ \mu^-$ can be used to make a
clear demarcation, not only between signals for RS and ADD gravitons, but
also between other kinds of new physics. This addresses an issue which has
not been discussed in any detail before and suggests an elegant and
easy-to-perform test, which uses cross-sections which are almost certain
to be measured when a high-energy linear collider actually goes into
operation.

\bigskip

\noindent {\large\bf Acknowledgments}

\noindent {\footnotesize This work constitutes part of the activities of
the Indian Linear Collider Working Group (ILCWG) under Project No.
SP/S2/K-01/2000-II, of the Department of Science \& Technology, Government
of India. The authors thank Sunanda~Banerjee, Debajyoti~Choudhury, Saurabh
D.~Rindani and K.~Sridhar for discussions. }

\bigskip

\centerline{\Large\bf Appendix: Decay width and Cross-section formulae}

\bigskip

\noindent
In this Appendix, we collect some useful formulae which are necessary to
evaluate the graviton cross-sections discussed in the text. 

\bigskip

\noindent $\S$ {\sl Graviton decay;} 
The dimensionless functions $\Delta^{(n)}_{P\bar P}$ required to calculate 
the graviton partial widths are listed below:
\footnotesize
\begin{eqnarray}
\Delta^{(n)}_{\gamma\gamma} & = & \frac{1}{5} \nonumber \\
\Delta^{(n)}_{gg} & = & \frac{8}{5} \nonumber \\
\Delta^{(n)}_{WW} & = & \frac{2}{5} \sqrt{1 - 4x_W^2}
\left( \frac{13}{12} + \frac{14}{39}r_W + \frac{4}{13}r_W^2 \right)
\theta(x_n - 2r_W) \nonumber \\
\Delta^{(n)}_{ZZ} & = & \frac{1}{5} \sqrt{1 - 4x_Z^2}
\left( \frac{13}{12} + \frac{14}{39}r_Z + \frac{4}{13}r_Z^2 \right)
\theta(x_n - 2r_Z) \nonumber \\
\Delta^{(n)}_{HH} & = & \frac{1}{30} \left(1 - 4x_H^2 \right)^{5/2}
\theta(x_n - 2r_H) \nonumber \\
\Delta^{(n)}_{\nu \bar\nu} & = & \frac{1}{10} \nonumber \\
\Delta^{(n)}_{\ell\bar\ell} & = & \frac{1}{10} \left(1 - 4x_\ell^2 \right)^{3/2}\left( 1 + \frac{8}{3}r_\ell \right) \theta(x_n - 2r_\ell) \nonumber \\
\Delta^{(n)}_{q\bar q} & = & \frac{3}{10} \left(1 - 4x_q^2 \right)^{3/2}
\left( 1 + \frac{8}{3}r_q \right) \theta(x_n - 2r_q)
\end{eqnarray}
\normalsize
where $r_P = m_P/m_0$ for every SM particle $P$. These formulae are
consistent with those presented in Ref.~\cite{Lykken}.
When using the above formulae we must remember to sum over three neutrino
flavours $\nu$, three charged leptons $\ell$ and six quarks $q$. The 
colour factor for quarks is already included.  

\bigskip

\noindent $\S$ {\sl Single photon production:} 
The amplitudes corresponding to the 9 different Feynman diagrams in Figure~2 
are given below. Helicities of the final states are not exhibited explicitly, 
as these will be summed over. We have used the Feynman rules and notations
given in Ref.~\cite{Lykken}.
\footnotesize
\begin{eqnarray}
M_1 & = & 
\frac{ieg^2}{16\cos^2\theta_W} \frac{1}{(k_1 - p_1)^2} 
\frac{1}{(p_2+p_3)^2 - M_Z^2 + iM_Z\Gamma_Z} \varepsilon_\mu^\ast(p_1)
\nonumber \\
&& \times \bar{v}(k_2,\lambda_2) 
\gamma^\alpha (c_V - \gamma_5) (\not{k_1} - \not{p_1}) \gamma^\mu 
u(k_1, \lambda_1)
~.~\bar{u}(p_2) \gamma_\alpha(1 - \gamma_5) v(p_3)
\\
M_2 & = & 
\frac{-ieg^2}{16\cos^2\theta_W} \frac{1}{(k_2 - p_1)^2} 
\frac{1}{(p_2+p_3)^2 - M_Z^2 + iM_Z\Gamma_Z} \varepsilon_\mu^\ast(p_1)
\nonumber \\
&& \times \bar{v}(k_2,\lambda_2) 
\gamma^\mu (\not{k_2} - \not{p_1}) \gamma^\alpha (c_V - \gamma_5) 
u(k_1, \lambda_1)
~.~\bar{u}(p_2) \gamma_\alpha(1 - \gamma_5) v(p_3)
\\
M_3 & = & 
\frac{-ieg^2}{8} \frac{1}{(k_1 - p_2)^2 - M_W^2} \frac{1}{(k_2 - p_3)^2 
- M_W^2} \nonumber \\
&& \left\{ (k_1 - k_2 - p_2 + p_3)^\mu \eta^{\nu\lambda}
      + (k_2 + p_1 - p_3)^\nu \eta^{\lambda\mu}
      - (k_1 + p_1 - p_2)^\lambda \eta^{\mu\nu} \right\} 
\varepsilon_\mu^\ast(p_1)
\nonumber \\
&& \times \bar{v}(k_2,\lambda_2) \gamma_\lambda (1 - \gamma_5) v(p_3)
~.~\bar{u}(p_2) \gamma_\nu(1 - \gamma_5) u(k_1, \lambda_1)
\\
M_4 & = & 
\frac{-ieg^2}{8} \frac{1}{(k_1 - p_1)^2} \frac{1}{(k_2 - p_3)^2 - M_W^2}
\varepsilon_\mu^\ast(p_1)
\nonumber \\
&& \times \bar{v}(k_2,\lambda_2) \gamma^\alpha (1 - \gamma_5) v(p_3)
~.~\bar{u}(p_2) \gamma_\alpha(1 - \gamma_5) (\not{k_1} - \not{p_1}) 
\gamma^\mu u(k_1, \lambda_1)
\\
M_5 & = & 
\frac{ieg^2}{8} \frac{1}{(k_2 - p_1)^2} \frac{1}{(k_1 - p_2)^2 - M_W^2}
\varepsilon_\mu^\ast(p_1)
\nonumber \\
&& \times \bar{v}(k_2,\lambda_2) \gamma^\mu (\not{k_2} - \not{p_1}) 
\gamma^\alpha (1 - \gamma_5) v(p_3)
~.~\bar{u}(p_2) \gamma_\alpha(1 - \gamma_5) u(k_1, \lambda_1)
\\
M_6 & = & 
\frac{ie}{128} \Lambda(Q^2) \frac{1}{(k_1 - p_1)^2} \varepsilon_\mu^\ast(p_1)
P_{\alpha\beta\rho\sigma}(Q)
\nonumber \\
&& \times \bar{v}(k_2,\lambda_2) 
\left\{ \gamma^\alpha (k_1 - k_2 - p_1)^\beta 
      + \gamma^\beta (k_1 - k_2 - p_1)^\alpha 
      - 2 \eta^{\alpha\beta} (\not{k_1} - \not{p_1}) \right\}
(\not{k_1} - \not{p_1}) \gamma^\mu u(k_1, \lambda_1)
\nonumber \\
&& \times \bar{u}(p_2) 
\left\{ \gamma^\rho (p_2 - p_3)^\sigma
      + \gamma^\sigma (p_2 - p_3)^\rho \right\} v(p_3)
\\
M_7 & = & 
\frac{-ie}{128} \Lambda(Q^2) \frac{1}{(k_2 - p_1)^2} 
\varepsilon_\mu^\ast(p_1) P_{\alpha\beta\rho\sigma}(Q)
\nonumber \\
&& \times \bar{v}(k_2,\lambda_2) \gamma^\mu (\not{k_2} - \not{p_1})
\left\{ \gamma^\alpha (k_1 - k_2 + p_1)^\beta
      + \gamma^\beta (k_1 - k_2 + p_1)^\alpha
      + 2 \eta^{\alpha\beta} (\not{k_2} - \not{p_1}) \right\}
u(k_1, \lambda_1)
\nonumber \\
&& \times \bar{u}(p_2)
\left\{ \gamma^\rho (p_2 - p_3)^\sigma
      + \gamma^\sigma (p_2 - p_3)^\rho \right\} v(p_3)
\\
M_8 & = & 
\frac{ie}{32} \Lambda(Q^2) \frac{1}{(k_1 + k_2)^2} \varepsilon_\mu^\ast(p_1)
P_{\alpha\beta\rho\sigma}(Q)
\left\{ (k_1 + k_2).p_1 C^{\alpha\beta\mu\nu} + D^{\alpha\beta\mu\nu} 
\right\}
\nonumber \\
&& \times \bar{v}(k_2,\lambda_2) \gamma_\nu u(k_1, \lambda_1)
~.~ \bar{u}(p_2)
\left\{ \gamma^\rho (p_2 - p_3)^\sigma
      + \gamma^\sigma (p_2 - p_3)^\rho \right\} v(p_3)
\\
M_9 & = & 
\frac{-ie}{64} \Lambda(Q^2) \varepsilon_\mu^\ast(p_1) 
P_{\alpha\beta\rho\sigma}(Q) 
\left\{ C^{\alpha\beta\mu\lambda} - \eta^{\alpha\beta} \eta^{\mu\lambda} 
\right\}
\nonumber \\
&& \times \bar{v}(k_2,\lambda_2) \gamma_\lambda u(k_1, \lambda_1)
~.~ \bar{u}(p_2)
\left\{ \gamma^\rho (p_2 - p_3)^\sigma
      + \gamma^\sigma (p_2 - p_3)^\rho \right\} v(p_3)
\end{eqnarray}
\normalsize
where $c_V = 1 - 4\sin^2\theta_W \simeq 0.074$. The sum over graviton
polarisations reduces, when the graviton couples to a conserved current
with massless fermions, to the simple form
\begin{equation}
P_{\mu\nu\rho\sigma}(Q) = 
\eta_{\mu\rho} \eta_{\nu\sigma}
+ \eta_{\nu\rho} \eta_{\mu\sigma}
\end{equation}
where $Q = p_2 + p_3$ is the momentum carried by the graviton propagator. 
The tensor couplings 
$C^{\alpha\beta\mu\nu}$ and $D^{\alpha\beta\mu\nu}$ are given by
\begin{eqnarray}
C^{\alpha\beta\mu\nu} & = &
\eta^{\alpha\mu} \eta^{\beta\nu} + \eta^{\alpha\nu} \eta^{\beta\mu}
- \eta^{\alpha\beta} \eta^{\mu\nu}
\\
D^{\alpha\beta\mu\nu} & = & 
        \eta^{\alpha\beta} p_1^\nu (k_1 + k_2)^\mu
-\left\{ \eta^{\alpha\nu} p_1^\beta (k_1 + k_2)^\mu
      + \eta^{\alpha\mu} p_1^\nu (k_1 + k_2)^\beta
      - \eta^{\mu\nu} p_1^\alpha (k_1 + k_2)^\beta \right\}
\nonumber \\
&& \hspace*{1.12in} -\left\{ \alpha \leftrightarrow \beta \right\}
\nonumber 
\end{eqnarray}
following Ref.~\cite{Lykken}. The function $\Lambda(Q^2)$ is discussed 
in the text (Section 2).


\end{document}